# An initial exploration of vicarious and in-scene calibration techniques for small unmanned aircraft systems

Baabak G. Mamaghani, Geoffrey V. Sasaki, Ryan J. Connal, Kevin Kha, Jackson S. Knappen, Ryan A. Hartzell, Evan D. Marcellus, Timothy D. Bauch, Nina G. Raqueño and Carl Salvaggio

Rochester Institute of Technology, College of Science, Chester F. Carlson Center for Imaging Science, 54 Lomb Memorial Drive, Rochester, NY 14623-5604, USA

## ABSTRACT

The use of small unmanned aircraft systems (sUAS) for applications in the field of precision agriculture has demonstrated the need to produce temporally consistent imagery to allow for quantitative comparisons. In order for these aerial images to be used to identify actual changes on the ground, conversion of raw digital count to reflectance, or to an atmospherically normalized space, needs to be carried out. This paper will describe an experiment that compares the use of reflectance calibration panels, for use with the empirical line method (ELM), against a newly proposed ratio of the target radiance and the downwelling radiance, to predict the reflectance of known targets in the scene. We propose that the use of an on-board downwelling light sensor (DLS) may provide the sUAS remote sensing practitioner with an approach that does not require the expensive and time consuming task of placing known reflectance standards in the scene. Three calibration methods were tested in this study: 2-Point ELM, 1-Point ELM, and At-altitude Radiance Ratio (AARR). Our study indicates that the traditional 2-Point ELM produces the lowest mean error in band effective reflectance factor, 0.0165. The 1-Point ELM and AARR produce mean errors of 0.0343 and 0.0287 respectively. A modeling of the proposed AARR approach indicates that the technique has the potential to perform better than the 2-Point ELM method, with a 0.0026 mean error in band effective reflectance factor, indicating that this newly proposed technique may prove to be a viable alternative with suitable on-board sensors.

## 1. INTRODUCTION

Recent innovation in remote aircraft design has enabled remote sensing practitioners to work at resolutions and under conditions originally thought to be impractical with traditional remote sensing platforms such as satellites and manned aircrafts. Known as sUAS (small unmanned aerial systems), these platforms have become widespread throughout the consumer market leading to an explosion of sUAS being used for untested remote sensing applications.[1] With a sUAS ready to fly at a moments notice, remotely sensed images are able to be captured at low altitudes (below 400ft) with atmospheric conditions ranging from ideal clear sunny days to overcast cloudy days. These images need to undergo proper calibration, otherwise incorrect results and conclusions will be drawn from them. With this newfound ease of collection, we must not forget the importance of the lessons learned from traditional remote sensing practices.

Remotely sensed images are affected by a variety of factors. These factors include, but are not limited to, date/time, geographic location, and atmospheric conditions. To mitigate these components and produce accurate results, calibration needs to take place. As an example, Figure 1 displays the need for calibration. Radiance images normalized by the exposure and gain, of the same scene on a cloudy and sunny day, do not appear the same, yet when converted to reflectance, the images appear invariant to illumination conditions. A traditional agricultural remote sensing metric, Normalized Difference Vegetation Index (NDVI),[2] calculated using both radiance and reflectance are presented in Figure 2.

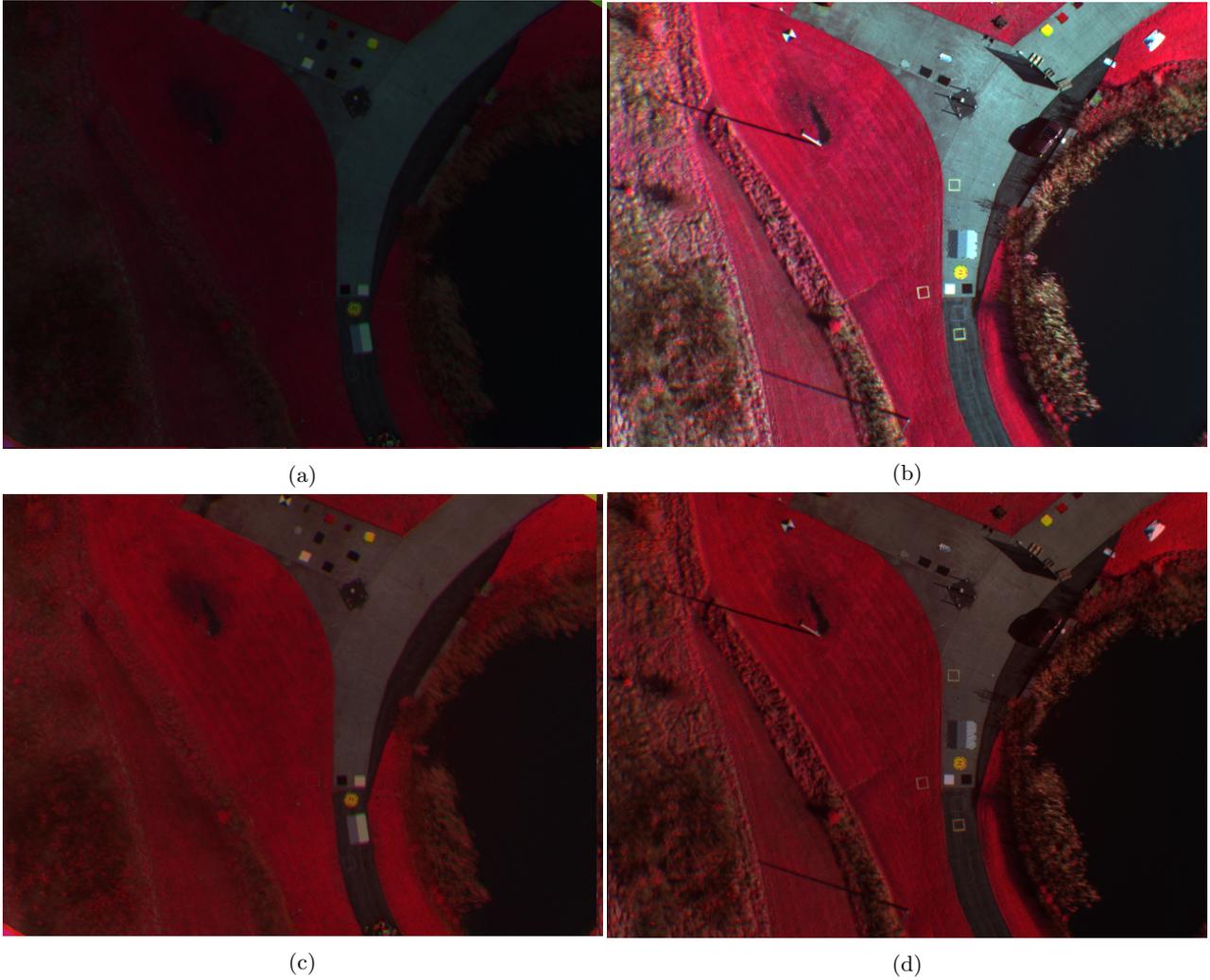

Figure 1: Color infrared (CIR) radiance imagery collected under (a) cloudy and (b) sunny conditions. Derived reflectance maps for the (c) cloudy and (d) sunny conditions are shown. The imagery was collected using a MicaSense RedEdge sensor for an altitude of 375ft on November 2, 2017 (cloud) and November 8, 2017 (sunny). Note the apparent change in brightness consistent with a cloudy atmosphere is removed when the image is converted to reflectance.

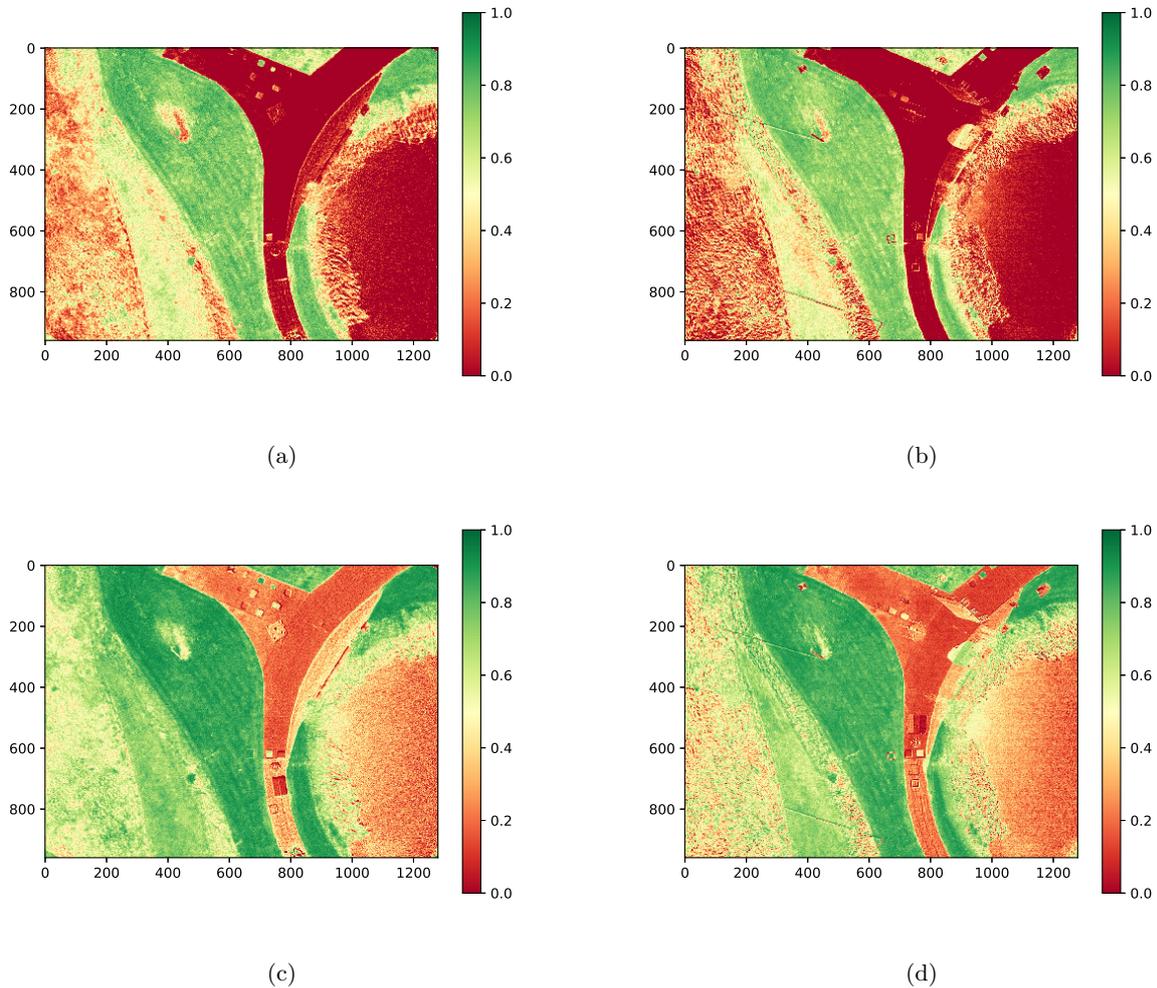

Figure 2: NDVI maps derived with radiance imagery under (a) cloudy and (b) sunny conditions. NDVI maps derived using reflectance maps for the (c) cloudy and (d) sunny conditions are shown. The imagery was collected using a MicaSense RedEdge sensor for an altitude of 375ft on November 2, 2017 (cloud) and November 8, 2017 (sunny). Note the difference in the radiance NDVI of the grass in the lower left corner of the NDVI maps which became consistent in the reflectance images.

The purpose of this study is to compare the accuracy of the widely used 2-Point and 1-Point ELM calibration techniques to a newly proposed At-altitude Radiance Ratio technique.

## 2. BACKGROUND

### 2.1 Empirical Line Method

The Empirical Line Method (ELM) is one of the most well known methods of converting radiance to reflectance for any remotely sensed image. By placing suitably sized calibration panels in the scene, a linear relationship can be determined between digital count and reflectance. The linear relationship can be determined if the reflectance of the calibration panel(s) is known at the time of data collection (to best account for directional effects). For results that best approximate the linear relationship, two panels are recommended (one bright and one dark) and have the properties of homogeneousness and are parallel to the ground.[3,4] In 2008, Baugh and Groeneveld empirically proved the ELM.[5]

Other researchers have implemented the ELM techniques differently. Wang et al.[6] showed results that the relationship of raw digital number and the natural log of the surface reflectance is linear. Furthermore, they state that the y-intercept is the minimum reflectance of the surface, and only one data point (panel) is needed to compute the digital number (digital count) to reflectance look up table. Ultimately, their results show that there is no statistically significant difference between the measured and predicted reflectance values.

## 2.2 MicaSense RedEdge

MicaSense offers remote sensing imagery solutions for agricultural use. Their RedEdge multispectral camera is a five band camera that captures multiple filtered images simultaneously. They offer a DLS (downwelling light sensor) that measures the downwelling irradiance for the same spectral bands that can be mounted on a sUAS platform. Our experiment utilizes this sensor to measure the lighting conditions for any given image and is integral to the proposed calibration technique. Table 1 displays the five MicaSense RedEdge spectral bands along with their respective wavelengths and bandwidth (FWHM).[7] A picture of the MicaSense RedEdge camera and the DLS can be seen in Figure 3.

Table 1: MicaSense RedEdge spectral bands with respective center wavelengths and bandwidth values.

| Band Name | Wavelengths [nm] | FWHM [nm] |
|---|---|---|
| Blue | 475 | 20 |
| Green | 560 | 20 |
| Red | 668 | 10 |
| Red Edge | 717 | 10 |
| Near IR | 840 | 40 |

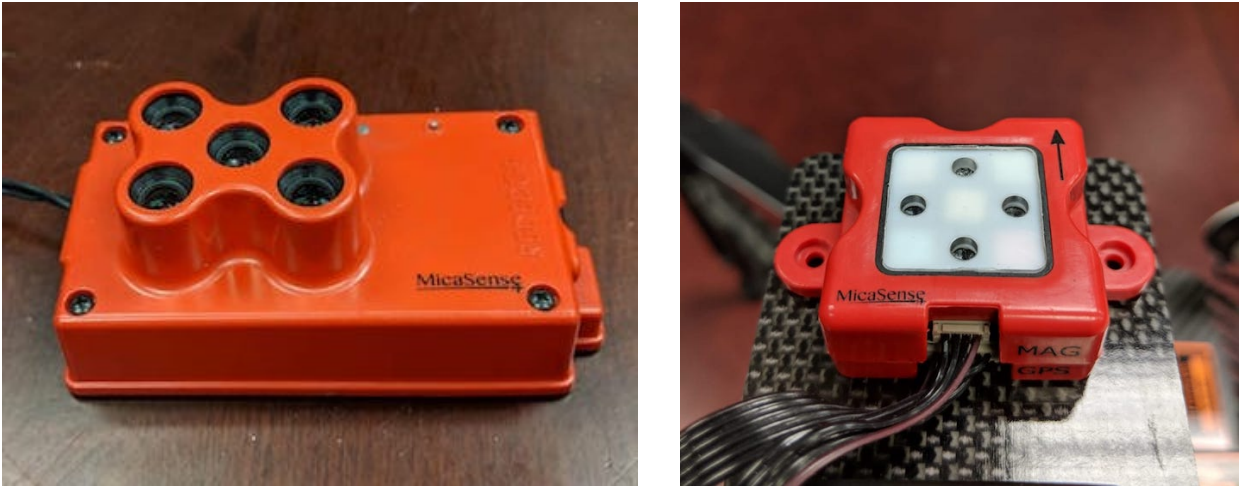

(a)        (b)

Figure 3: MicaSense RedEdge (a) camera, and (b) downwelling light sensor

The metadata of each captured image contains fixed values, for each band, to convert the digital count values to radiance ($W/m^2/nm/sr$). These values include a vignette function, radiometric calibration coefficients, sensor gain, exposure time and black level values.

To model the MicaSense RedEdge's output correctly, the relative spectral response (RSR) functions of the camera was computed using spectral power data measured using a Newport Model 74004-1 monochromator. Figure 4 displays these RSR functions for the five spectral channels of the MicaSense RedEdge camera.

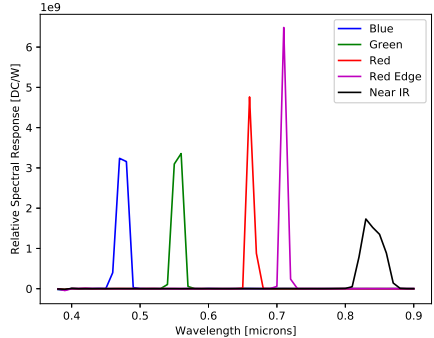

Figure 4: MicaSense relative spectral response functions (sampling interval of 10nm and bandpass of 1nm).

## 2.3 Sensor Reaching Radiance

The spectral radiance reaching a sensor[8] can be expressed as:

$$L_s(\lambda) = \left[\frac{E'_{solar}(\lambda)}{\pi}cos(\sigma')\tau_1(\lambda)\rho(\lambda) + L_{\downarrow solar}(\lambda)\rho_d(\lambda)\right]\tau_2(\lambda) + L_{\uparrow solar}(\lambda) + L_a \qquad (1)$$

where, $E'_{solar}(\lambda)$ is the spectral exoatmospheric solar irradiance, $\sigma'$ is the solar elevation angle, $\tau_1(\lambda)$ is the spectral transmission from space to the target, $\tau_2(\lambda)$ is the spectral transmission from target to sensor, $\rho(\lambda)$ is the spectral directional reflectance function for the target, $\rho_d(\lambda)$ is the spectral diffuse reflectance for the target, $L_{\downarrow solar}(\lambda)$ is the solar scattered downwelling sky radiance propagating on to the target, $L_{\uparrow solar}(\lambda)$ is the solar scattered spectral path radiance generated in the path between target and the sensor, and $L_a(\lambda)$ is the scattered photons from background objects near the target of interest.

Band effective radiance, $L_{s,i}$, can be computed by integrating the spectral radiance given by Equation 1 over the spectral bandpass defined by the RSR functions for each of the five bands of the MicaSense RedEdge sensor using

$$L_{s,i} = \frac{\int_\lambda L_s(\lambda) RSR_i(\lambda) d\lambda}{\int_\lambda RSR_i(\lambda) d\lambda} \text{ for } i = [1, 5] \qquad (2)$$

where $RSR_i(\lambda)$ is the relative spectral response for the $i^{th}$ band.

## 3. METHODOLOGY

The first method is a 2-Point ELM which applied a look up table (LUT), computed from two in scene calibration panels, to convert digital count to reflectance. The second method is a 1-Point ELM that applied a radiance to reflectance factor to every pixel in the image (after converting the image from digital count to radiance). Both ELM methods were applied to five images that were captured during a sUAS flight on November 8, 2017 at 375ft.

Afterwards, a newly proposed technique, which we refer to as At-altitude Radiance Ratio (AARR), is tested on the same five images. This approach utilized the recorded DLS spectral irradiance values and divides the converted radiance image by the converted DLS radiance values. The reflectance factor errors from all three of these methods were computed and compared against one another.

### 3.1 2-Point Empirical Line Method (2-Point ELM)

The first method tested was the 2-Point ELM. Raw digital counts were normalized by their gain and exposure time. The normalized digital count in each band, $DC_{norm,i}$, is given by

$$DC_{norm,i} = DC_{raw,i} \left(\frac{t_{min}}{t_i}\right) \left(\frac{g_{min}}{g_i}\right) \left(\frac{2^n - 1}{2^m - 1}\right) \text{ for } i = [1, 5] \quad (3)$$

where $DC_{raw,i}$ denotes the raw digital count of each pixel, $t_{min}$ denotes the shortest exposure time the sensor system is capable of (0.066ms), and $t$ is the exposure time of the current frame. Similarly, $g_{min}$ represents the minimum gain, or ISO, possible (1x or 100) and $g$ is the gain used in capturing the current frame. $n$ is the bit depth of the desired normalized digital count dynamic range and $m$ is the sensor bit depth.

A LUT is calculated for each band by relating the normalized pixel value of a calibration panel to the measured reflectance of that panel. One bright and one dark reflectance calibration panel were used in the scene. From these two points, ELM was conducted by finding the slope and bias which converts normalized pixel value to reflectance $\rho_i$ described by

$$\rho_i = b_{1,i} DC_{norm,i} + b_{0,i} \quad (4)$$

$$b_{1,i} = \frac{\rho_{bright,i} - \rho_{dark,i}}{DC_{norm,bright,i} - DC_{norm,dark,i}} \quad (5)$$

$$b_{0,i} = \rho_{bright,i} - b_{1,i} DC_{norm,bright,i} \quad (6)$$

where $b_{1,i}$ is the slope, $\rho_{bright,i}$ is the band effective reflectance of the bright panel for band $i$, $\rho_{dark,i}$ is the band effective reflectance of the dark panel for band $i$, $DC_{norm,bright,i}$ is the normalized digital count of the bright panel for band $i$, and $DC_{norm,dark,i}$ is the normalized digital count of the dark panel for band $i$.

A unique series of spectral LUTs is only valid for a single illumination level. Illumination conditions can vary greatly during a single flight. By producing several LUTs, under varying cloud conditions, illumination variation can be handled. For this study, images were selected where the two calibration panels were located nadir to the sensor (within a 10° radius of the image center), and a series of spectral LUTs was produced for each of those images. These LUTs were paired with the DLS measurements (a five element array) recorded with their respective image.

When converting an image to reflectance, The LUT with the smallest Euclidean distance between it's paired DLS measurement and the DLS measurement of the current image is applied to the image. Euclidean distance is described by

$$d(DLS_{img}, DLS_{LUT_k}) = \sqrt{\sum_{i=1}^{5}(DLS_{img,i} - DLS_{LUT_k,i})^2} \text{ for } k = [1, K] \quad (7)$$

where $d(DLS_{img}, DLS_{LUT_k})$ is the Euclidean distance between the DLS measurement of the image being converted to reflectance, $DLS_{img}$, and the DLS measurement of the current LUT, $DLS_{LUT_k}$. $K$ is the number of LUTs for the current sUAS flight, and $i$ denotes the spectral band of DLS measurement. This ultimately results in a reflectance image that has been adjusted for variation in illumination.

## 3.2 1-Point Empirical Line Method (1-Point ELM)

The second method tested was a 1-point ELM using code provided by MicaSense.[9] Here, vignetting correction, dark level bias, and radiometric calibration are also accounted for in the conversion of digital counts to radiance, $L_{s,i}$ as

$$I(x,y) = \left(\frac{a_{1,i}}{g_i}\right)\left(\frac{p - p_{BL,i}}{t_i + a_{2,i}y - a_{3,i}t_i y}\right) \text{ for } i = [1, 5] \tag{8}$$

where $a_{1,i}, a_{2,i}$ and $a_{3,i}$ are the radiometric calibration coefficients, $g_i$ is the sensor gain, $p$ and $p_{BL,i}$ are the normalized raw pixel value and normalized black level value, respectively, and $t_i$ is the exposure time of the image.

The vignette model applies a light sensitivity fall-off correction to all pixels in the image. Pixels that fall further from the vignette center have a larger factor applied. Equations 9-11 show how the vignette model is applied to each pixel in the image.

$$r = \sqrt{(x - c_x)^2 + (y - c_y)^2} \tag{9}$$

$$k = 1 + k_0 r + k_1 r^2 + k_2 r^3 + k_3 r^4 + k_4 r^5 + k_5 r^6 \tag{10}$$

$$V(x,y) = \frac{1}{k} \tag{11}$$

$$L_{s,i} = \frac{I(x,y)}{k} \tag{12}$$

where $c_x$ and $c_y$ represent the vignette center, $k_0$ through $k_5$ are polynomial correction coefficients extracted from the image metadata, $I(x,y)$ is the normalized pixel intensity from Equation 8. At this point, the radiance $L_{s,i}$ can be found at any pixel within an image, and the ELM method described in the previous section can then be applied in the same fashion, except the dark reflectance point is replaced by the origin ($L_{s,i} = 0$ and $\rho_i = 0$). This means the slope computed by ELM is a radiance to reflectance factor.

## 3.3 At-altitude Radiance Ratio (AARR)

To test the efficacy of the MicaSense RedEdge's DLS, a new method for converting raw digital counts into reflectance was examined. The images were converted to radiance the same way as the 1-Point ELM method (vignetting correction, radiometric calibration, dark level bias, gain and exposure time normalization). The DLS spectral irradiance was extracted from the image's metadata, and converted to radiance (divided by $\pi$). These two values were ratioed using Equation 13 to produce reflectance images for each spectral band. This AARR reflectance equation can be rewritten using the terminology from the sensor reaching radiance equation, as shown in Equation 14.

$$\rho_i = \frac{\text{Sensor Radiance}_i}{\text{DLS Radiance}_i} \text{ for } i = [1, 5] \tag{13}$$

$$\rho_i = \frac{L_{s,i}(\lambda)}{\frac{E'_{solar}(\lambda)}{\pi}cos(\sigma')\tau_1(\lambda) + L_{\downarrow solar,i}(\lambda)} \tag{14}$$

Others have applied similar techniques for conversion to reflectance. Lekki et al obtained reflectance by taking the ratio of the water leaving radiance and the downwelling irradiance,[10] while Ortiz et al took the ratio of the HSI2 (John Glenn Research Centers Hyperspectral Imager) radiance and the upward looking Analytical Spectral Devices (ASD) FieldSpec that they had fixed on their aircraft, which was flying at traditional remote sensing altitudes.[11] Our method differs because our sUAS was flying at a max height of 375ft, and the MicaSense RedEdge DLS recorded single irradiance values per band, not a full downwelling irradiance spectra.

## 3.4 Modeled At-altitude Radiance Ratio (M-AARR)

To demonstrate the theoretically achievable results by the AARR method, simulations were accomplished using Spectral Science Incorporated's moderate resolution atmospheric transmission (MODTRAN) code, specifically, MODTRAN4. MODTRAN is designed to model atmospheric propagation over the wavelength range for $0.2\mu m$ to $100\mu m$.[12] With the ability to model any combination of atmosphere, day, time or location, MODTRAN gives users the ability to simulate countless scenarios. For the purposes of this paper, the focus was primarily on matching the atmospheric conditions of November 8, 2017 in Henrietta, NY, USA (43.041099 N, 77.698343 W).

To run MODTRAN, an input file (named 'tape5') with all the simulation parameters needs to be provided. Each MODTRAN run solves the radiative transfer equation and provides accurate modeling of the atmosphere given the inputted parameters.[12] Therefore, a script was written to modify the input file based on the given parameters. Some of modified parameters were: target, background and sensor altitude. A full list of selected parameters can be seen in Appendix B. In addition, MODTRAN allows users to input their own reflectance curves to help generate models of measured reflectances. The mean ground reference reflectance curves for three targets (grass, concrete, and asphalt) was computed and added to the spectral albedo (named 'spec_alb.dat') file, which can be called by MODTRAN for modeling particular types of targets. This allowed for the modeling of the sensor reaching radiance of the targets that were measured in the field.

Once finished, MODTRAN provides an output file (named 'tape7.scn'), which contains spectral transmittance, radiance and irradiance components. Some of the computed variables are ground reflected radiance (GRND_RFLT), direct solar radiance (DRCT_RFLT), total radiance seen by the sensor (TOTAL_RAD) and the solar scattered radiance (SOL_SCAT). For the purposes of this study, we used both the total radiance, and the ground reflected radiance variables. The total radiance is represented fully in Equation 1 as $L_s(\lambda)$ and the ground reflected radiance is equivalent to the denominator of Equation 14.

To model the data that was seen by the camera, as well as the DLS, two separate MODTRAN runs were executed for every combination of simulation parameters. The first run was a measurement of the sensor reaching radiance. For this, the camera was pointed directly down at the target on the ground. The total radiance from this simulation was used as the sensor reaching radiance. The second run was to simulate the DLS downwelling radiance. This was important to model correctly, because depending on the atmospheric conditions, the sunlight can be between 10 and 100 times the magnitude of the skylight.[13] To model the DLS downwelling radiance, a constant 100% reflector was placed at the original sensor altitude, and the sensor altitude was raised by 1 meter. The ground reflected radiance of the 100% reflector was recorded and used as the DLS downwelling radiance. By simulating the reflected radiance of a constant 100% reflector, at the original sensor altitude, this value is equivalent to the downwelling radiance onto the DLS, because all of the sun and skylight are being reflected back up to the new sensor position (1m above).

To compute the at-altitude reflectance, the total radiance from MODTRAN run one was divided by the ground reflected radiance from MODTRAN run two. This can be seen in Equation 15, using the MODTRAN terminology, and in Equation 16, using the sensor reaching radiance terminology.

$$\rho(\lambda) = \frac{TOTAL\_RAD_1}{GRND\_RFLT_2} \quad (15)$$

$$\rho(\lambda) = \frac{L_s(\lambda)}{\frac{E'_{solar}(\lambda)}{\pi}cos(\sigma')\tau_1(\lambda) + L_{\downarrow solar}(\lambda)} \quad (16)$$

where $\rho(\lambda)$ is the reflectance of the target, $TOTAL\_RAD_1$ is the total radiance outputted from MODTRAN in the first run and $GRND\_RFLT_2$ is the ground reflected radiance outputted from MODTRAN in the second run. These two simulations are visualized in Figure 5 below.

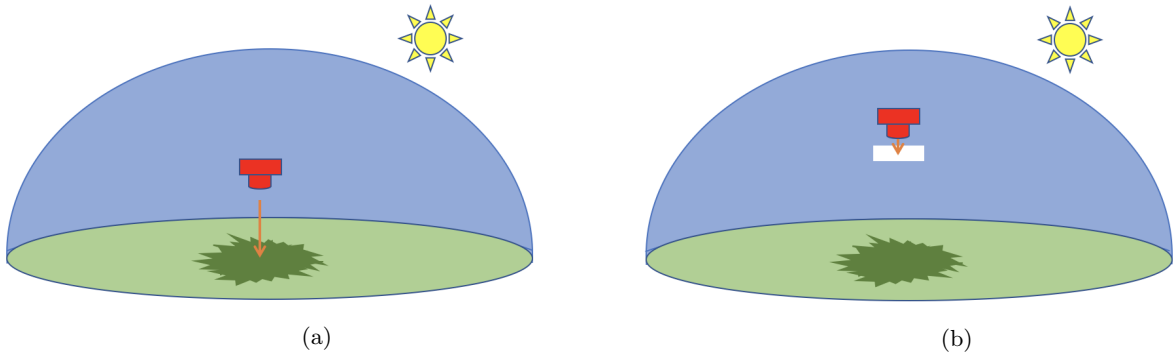

(a)                                                    (b)

Figure 5: (a) Sensor reaching radiance, and (b) downwelling radiance simulation diagrams.

## 3.5 Data Collection

For data collection, permission was given to fly our sUAS at the Henrietta Fire District Training Center (Station Number 6, 43.041099°N, 77.698343°W). In the scene, two sets of calibration panels were deployed (one set in direct sunlight and the other in shadow), and two sets of colored felts (red, green, blue and brown) were deployed, among other in scene targets. The tower casts a large shadow onto the ground as well, which will be helpful later when shadow detection and reflectance calibration of shadowed pixels is attempted in a follow up study. An example of a raw digital count orthomosaiced image from these collects is shown in Figure 6. The small unmanned aircraft system used for data collection was a DJI Matrice 100 quadcopter, which can be seen in Figure 7.

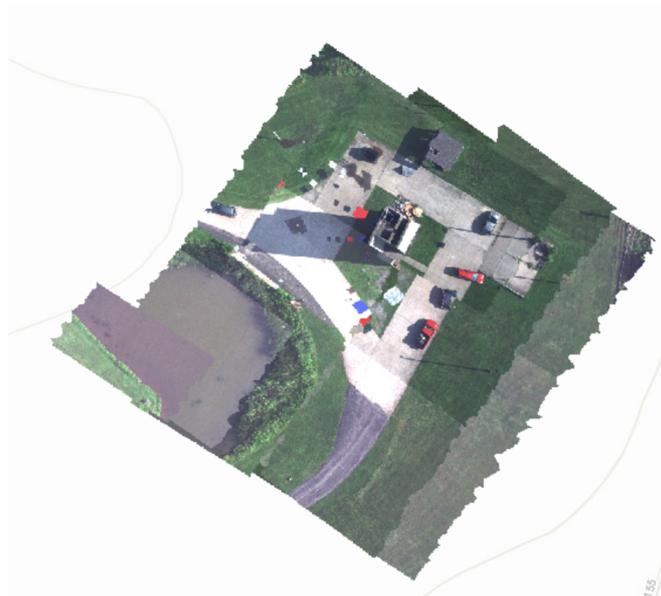

Figure 6: Henrietta Fire District Training Center orthomosaiced image. Calibration panels and test felts are distributed throughout the scene (in direct sunlight and shadow regions).

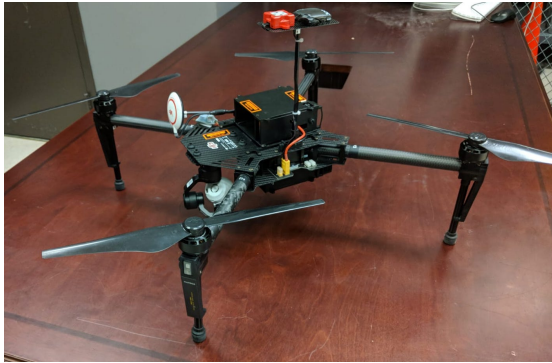
Figure 7: DJI Matrice 100 quadcopter used for data collection

Spectra Vista Corporation's (SVC) HR-1024i spectroradiometer was utilized for collecting ground reference data. With a spectral range of 350-2500nm, and use of 100% linear array detectors, the HR-1024i provides excellent wavelength stability, and produces highly accurate spectra.[14] To ensure the highest accuracy possible, a reference spectra was captured of a Spectralon panel before each measurement was taken. The spectra used for ground reference measurements can be found in Appendix A. Figure 8 shows an example spectra measurement in the field.

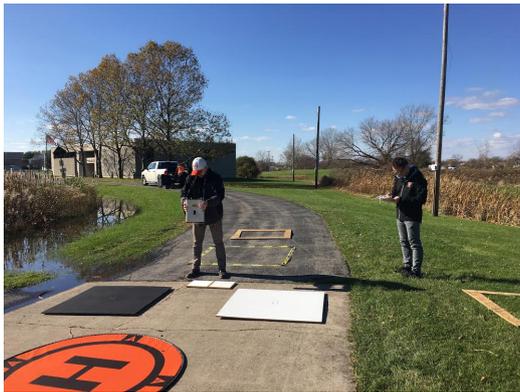
Figure 8: Example spectra collection in the field. Note that collector's shoulder is towards the sun. Measurements account for non-uniformity in the target by averaging an area of collection.

## 4. RESULTS

A table is presented with the band integrated ground reference reflectance factors for the three ground targets (grass, asphalt and concrete).

Table 2: Band integrated ground reference reflectance factors for in-scene targets derived from in scene measurements of observed targets using a spectra vista corporation spectroradiometer.

| Band [nm] | Grass | Asphalt | Concrete |
|---|---|---|---|
| Blue [475] | 0.0250 | 0.1030 | 0.1852 |
| Green [560] | 0.0763 | 0.1142 | 0.2332 |
| Red [668] | 0.0367 | 0.1215 | 0.2620 |
| Red Edge [717] | 0.1604 | 0.1236 | 0.2731 |
| NIR [840] | 0.4995 | 0.1299 | 0.2950 |

After all five sUAS images were converted to reflectance using 2-Point ELM, 1-Point ELM, and AARR, the computed reflectance factors of the three targets (grass, asphalt, and concrete) were compared against the band integrated ground reference reflectance factors. The average error and standard deviations, across the five images, were computed. Figures 9-11 display the computed statistics.

## 4.1 ELM and AARR

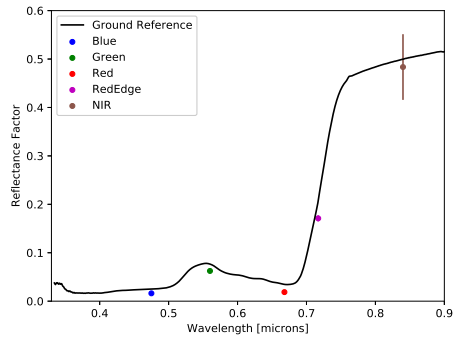

(a)

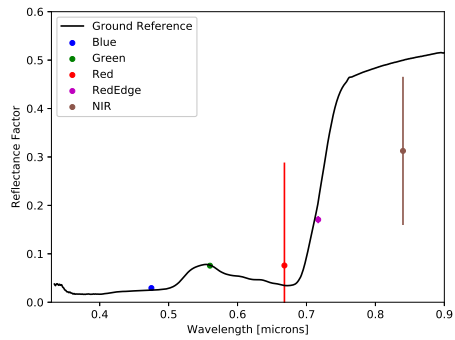

(b)

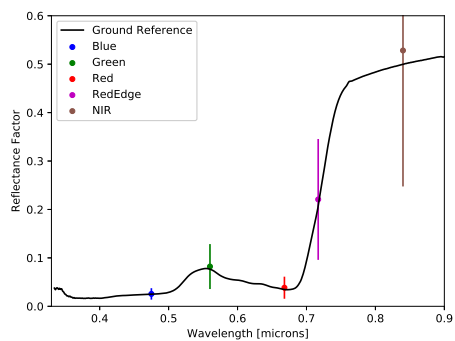

(c)

Figure 9: (a) 2-Point ELM, (b) 1-Point ELM, and (c) AARR applied to a grass target. Black curves are ground reference grass reflectance. Mean reflectance factor errors are plotted as dots with two standard deviation error bars shown.

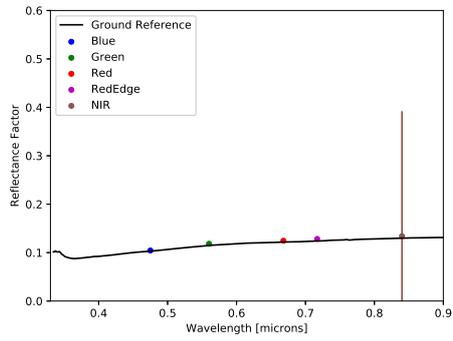

(a)

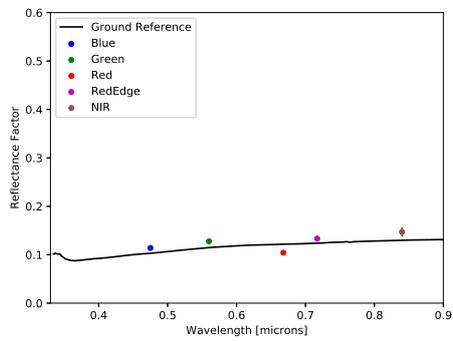

(b)

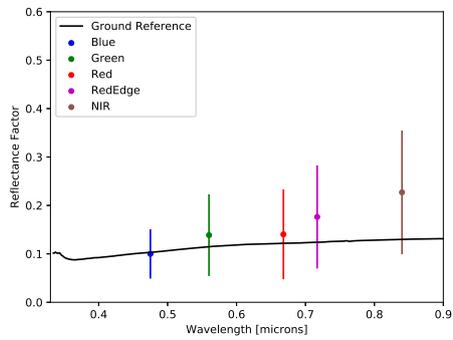

(c)

Figure 10: (a) 2-Point ELM, (b) 1-Point ELM, and (c) AARR applied to a asphalt target. Black curves are ground reference asphalt reflectance. Mean reflectance factor errors are plotted as dots with two standard deviation error bars shown.

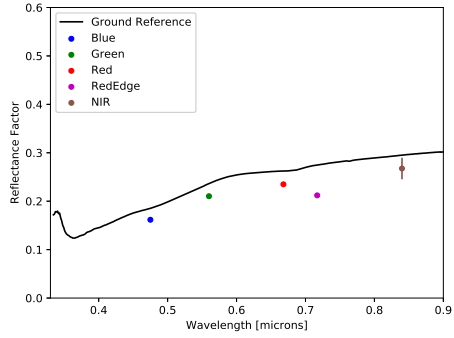

(a)

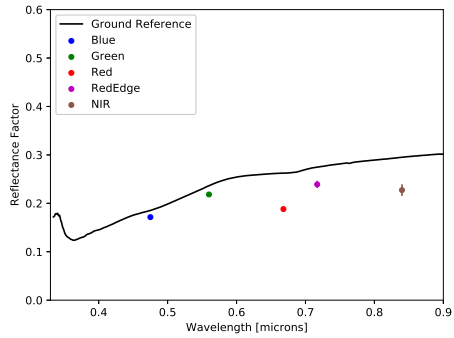

(b)

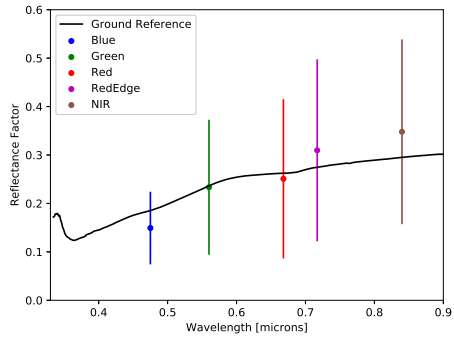

(c)

Figure 11: (a) 2-Point ELM, (b) 1-Point ELM, and (c) AARR applied to a concrete target. Black curves are ground reference concrete reflectance. Mean reflectance factor errors are plotted as dots with two standard deviation error bars shown.

### 4.2 M-AARR

For the purposes of this preliminary study, the only model simulated was designed to mimic the November 8, 2017 atmospheric conditions in Henrietta, NY. The model parameters for this simulation can be seen in Appendix B. Below are a set of figures that modeled the sensor reaching radiances, the downwelling radiances, and target reflectances vs the sensor altitudes. The six altitudes selected for these simulations were: 2, 150, 225, 300, 375, and 5000ft. 150, 225, 300 and 375ft were all selected because those were the heights of the sUAS data collections at the Henrietta Fire Training Center. 2ft was selected to test the M-AARR technique against the ground reference (these two should have been more or less identical), and 5000ft was simulated to show how the AARR method performs for traditional remote sensing.

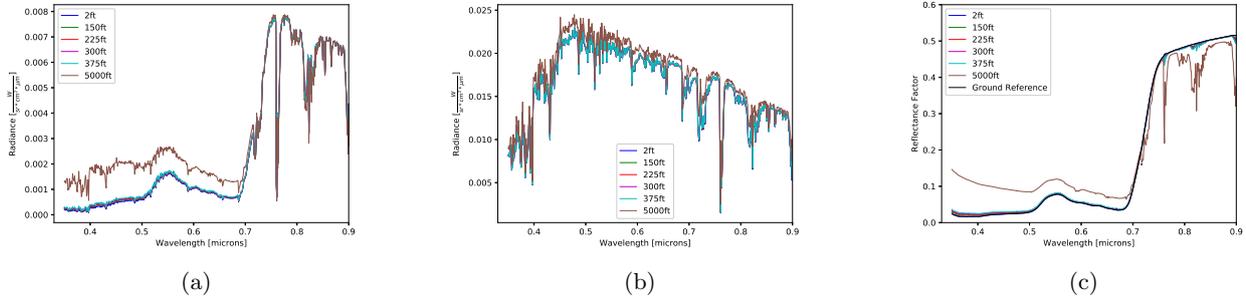

(a) (b) (c)

Figure 12: (a) Sensor reaching radiance, (b) downwelling radiance, and (c) reflectance of grass from MODTRAN modeling for six sensor altitudes. Scattered dots in the reflectance figure are the band integrated reflectances computed for all heights and the ground reference curves.

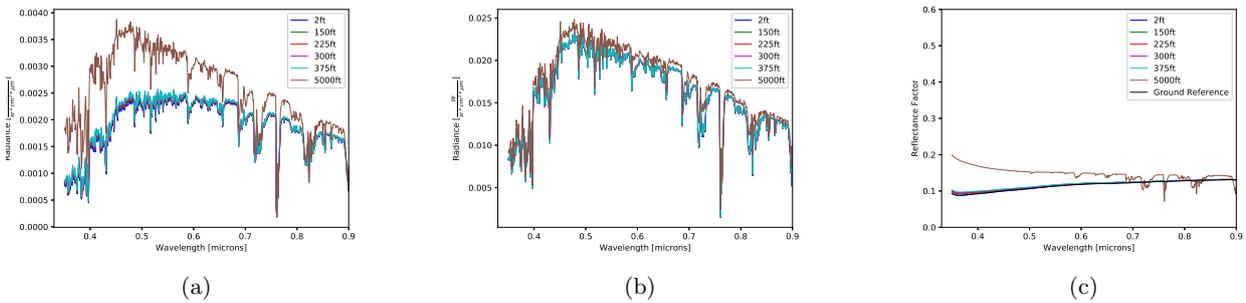

(a) (b) (c)

Figure 13: (a) Sensor reaching radiance, (b) downwelling radiance, and (c) reflectance of asphalt from MODTRAN modeling for six sensor altitudes. Scattered dots in the reflectance figure are the band integrated reflectances computed for all heights and the ground reference curves.

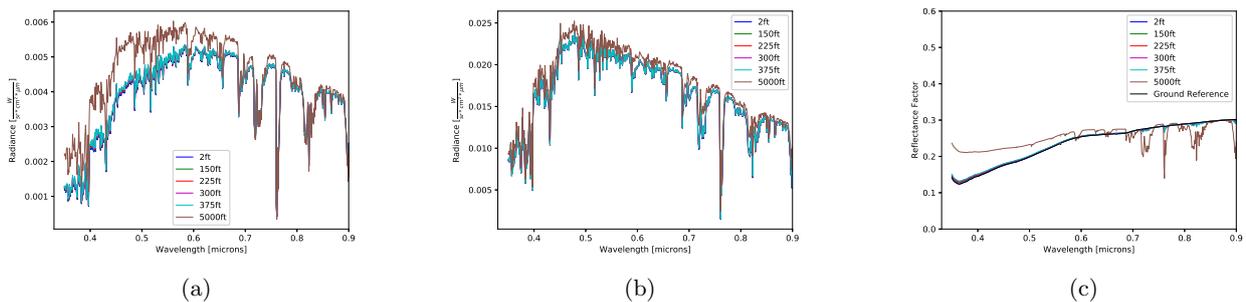

(a) (b) (c)

Figure 14: (a) Sensor reaching radiance, (b) downwelling radiance, and (c) reflectance of concrete from MODTRAN modeling for six sensor altitudes. Scattered dots in the reflectance figure are the band integrated reflectances computed for all heights and the ground reference curves.

Table 3: Reflectance factor errors for grass target. M-AARR are single point errors between simulation and ground reference values. 2-Point ELM, 1-Point ELM and AARR are averaged errors from five sUAS images. All values are computed from sUAS flights of 375ft. Positive means are overestimations of the reflectance factor, while negative means are underestimations.

|          | 2-Point ELM |         | 1-Point ELM |         | AARR    |         | M-AARR  |
|----------|-------------|---------|-------------|---------|---------|---------|---------|
| Channel  | Mean        | St. Dev | Mean        | St. Dev | Mean    | St. Dev | -       |
| Blue     | -0.0088     | 0.0009  | 0.0043      | 0.0007  | 0.0006  | 0.0059  | 0.0056  |
| Green    | -0.0139     | 0.0008  | -0.0008     | 0.001   | 0.0058  | 0.0232  | 0.0037  |
| Red      | -0.0180     | 0.0014  | 0.0392      | 0.1063  | 0.0017  | 0.0114  | 0.0029  |
| Red Edge | 0.0106      | 0.0031  | 0.0102      | 0.0037  | 0.0603  | 0.0624  | 0.0006  |
| NIR      | -0.0160     | 0.0339  | -0.1870     | 0.0765  | 0.0288  | 0.1404  | -0.0042 |

Table 4: Reflectance factor errors for asphalt target. M-AARR are single point errors between simulation and ground reference values. 2-Point ELM, 1-Point ELM and AARR are averaged errors from five sUAS images. All values are computed from sUAS flights of 375ft. Positive means are overestimations of the reflectance factor, while negative means are underestimations.

|          | 2-Point ELM |         | 1-Point ELM |         | AARR    |         | M-AARR  |
|----------|-------------|---------|-------------|---------|---------|---------|---------|
| Channel  | Mean        | St. Dev | Mean        | St. Dev | Mean    | St. Dev | -       |
| Blue     | 0.0014      | 0.0009  | 0.0110      | 0.0013  | -0.0031 | 0.0255  | 0.0048  |
| Green    | 0.0041      | 0.0014  | 0.0135      | 0.0011  | 0.0244  | 0.0422  | 0.0034  |
| Red      | 0.0031      | 0.0006  | -0.0173     | 0.0011  | 0.0188  | 0.0464  | 0.0022  |
| Red Edge | 0.0048      | 0.0019  | 0.0097      | 0.0029  | 0.0526  | 0.0532  | 0.0012  |
| NIR      | 0.0042      | 0.1290  | 0.0172      | 0.0048  | 0.0971  | 0.0639  | 0.0003  |

Table 5: Reflectance factor errors for concrete target. M-AARR are single point errors between simulation and ground reference values. 2-Point ELM, 1-Point ELM and AARR are averaged errors from five sUAS images. All values are computed from sUAS flights of 375ft. Positive means are overestimations of the reflectance factor, while negative means are underestimations.

|          | 2-Point ELM |         | 1-Point ELM |         | AARR    |         | M-AARR  |
|----------|-------------|---------|-------------|---------|---------|---------|---------|
| Channel  | Mean        | St. Dev | Mean        | St. Dev | Mean    | St. Dev | -       |
| Blue     | -0.0235     | 0.0016  | -0.0137     | 0.0013  | -0.036  | 0.0375  | 0.0039  |
| Green    | -0.0229     | 0.0013  | -0.0149     | 0.0016  | 0.0002  | 0.0697  | 0.0025  |
| Red      | -0.0271     | 0.0020  | -0.0739     | 0.0019  | -0.0111 | 0.0822  | 0.0011  |
| Red Edge | -0.0610     | 0.0014  | -0.0340     | 0.0037  | 0.0366  | 0.0939  | -0.0007 |
| NIR      | -0.0275     | 0.0111  | -0.0677     | 0.006   | 0.0530  | 0.0954  | -0.0017 |

For comparison purposes, the average absolute reflectance factor error across all channels and targets for each method was computed. 2-Point ELM, 1-Point ELM, AARR, and M-AARR produced 0.0165, 0.0343, 0.0287, and 0.0026 reflectance factor error, respectively.

While the standard deviations are much higher in the AARR, this is to be expected because of the metadata variables that were utilized for this technique. Among these metadata variables were the gain and exposure time which are discrete values that are set, for each band of the image, automatically, as the images are captured from the RedEdge camera. The gain can only be one of four values, while the exposure time can be one of 21 values. If these variables are not quantized, the results would be more representative of the conditions they were captured under and therefore, more accurate. ELM has a much lower standard deviation because the calibration panels used to compute the point slope formulas also underwent the same exposure and gain correction as the targets in the scene.

Overall, 2-Point ELM performed the best across all five bands and three targets. While AARR did not perform as well as the 2-Point ELM, it outperformed the 1-Point ELM. In addition, the M-AARR results showed that with the proper setup and instrumentation, AARR reflectance factor error could be lowered significantly. Therefore, given the proper instrumentation, sUAS imagery (to 400ft) can produce very reliable reflectance maps without the use of calibration panels.

## 5. FUTURE WORK

The initial study was only accomplished for a small set of parameters. The atmosphere, visibility, day and time were all purposefully selected to best model the data that was collected with our sUAS. We plan to expand the modeling section of our study to assess the feasibility of implementing AARR technique under various conditions, which include:

- **Model Atmosphere:** Tropical, Mid-Latitude Winter, and 1976 US Standard
- **Visibility:** 5, 10, 15, and 23km
- **Day Number:** March 20th, June 20th, September 22nd, and December 21st
- **Time:** 08:40, 10:20, 12:00, 13:50 and 15:40 local time.

This upcoming in-depth study will be aimed at providing an indication of quality of performance under various atmospheric conditions. This will provide sUAS users with an understanding of the reflectance accuracy they could produce without in-scene calibration panels. All of these simulations will be modeled using MODTRAN and analyzed using the M-AARR technique.

In addition to the extra modeling that will be accomplished, all the captured images from the 12 sUAS flights at the Henrietta Fire District Training Center will be analyzed using 2-Point ELM, 1-Point ELM and AARR. This will provide us with more reliable reflectance factor error statistics, as the averages will be across hundreds of images.

Finally, speaking with instrumentation manufacturers could help them in producing better cameras and downwelling light sensors which in turn would produce more reliable data for all sUAS users.

## 6. CONCLUSIONS

While 2-Point ELM produced the lowest mean error in band effective reflectance factor (0.0165), this preliminary study has shown that the AARR method is a reliable technique for producing reflectance imagery from sUAS captured images (0.0287 mean error). Under the assumption that the in-depth study produces similar results, for a wide range of atmospheric conditions, sUAS users will not need to place calibration panels or conduct ELM to convert their images to reflectance. This will save the sUAS remote sensing practitioner significant time during ground collects as they will not need to place calibration panels or collect ground reference spectra. In addition, sUAS users will also save time during data processing, because they would not have to go through the burden of creating and applying LUTs to convert the images to reflectance.

# APPENDIX A. REFLECTANCE SPECTRA

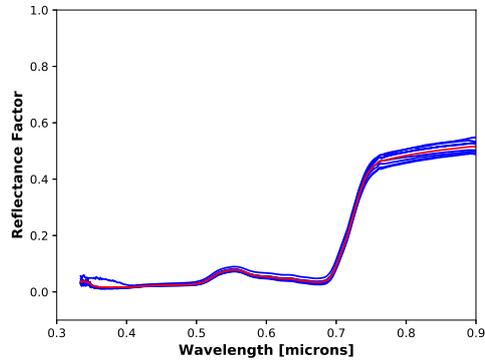

Figure 15: Collected grass spectra during sUAS flights. Blue spectra are single measurements while the red spectra is the average measurement across three days of data collection (November 2, 8, and 9, 2017).

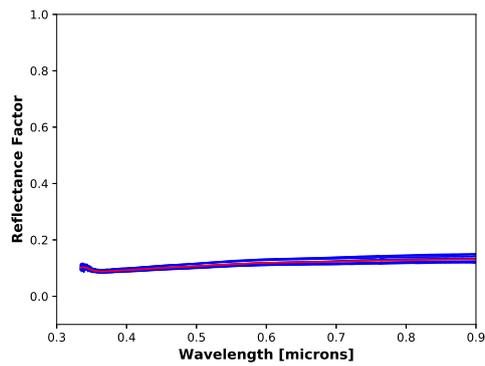

Figure 16: Collected asphalt spectra during sUAS flights. Blue spectra are single measurements while the red spectra is the average measurement across three days of data collection (November 2, 8, and 9, 2017).

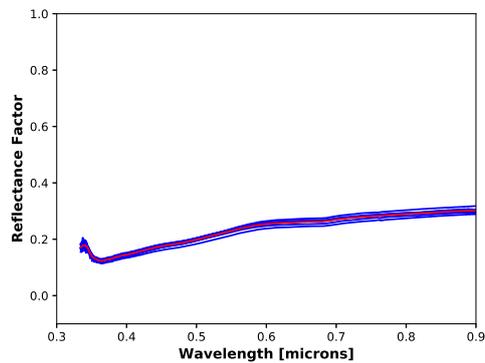

Figure 17: Collected concrete spectra during sUAS flights. Blue spectra are single measurements while the red spectra is the average measurement across three days of data collection (November 2, 8, and 9, 2017).

# APPENDIX B. MODTRAN PARAMETERS

Table 6: Description of the MODTRAN variables and the values utilized for simulation.[15]

| Description | MODTRAN Variable | Value |
|---|---|---|
| **Model Atmosphere** | MODEL | 2 (Mid-Latitude Summer) |
| **Path Type** | ITYPE | 2 (Slant or Vertical Path Between Two Altitudes) |
| **Surface Albedo** | SURREF | 'LAMBER' |
| **Surface Temperature [K]** | AATEMP | 303 |
| **Target** | CSLAB | Grass, Concrete, Asphalt, 100% Constant (from 'spec_alb.dat') |
| **Background** | CSLAB | Grass, Concrete, Asphalt (from 'spec_alb.dat') |
| **Visibility [km]** | VIS | 15.0 |
| **Ground Altitude [km]** | GNDALT | 0.168 |
| **Sensor Altitude [km]** | H1 | 0.169, 0.214, 0.237, 0.259, 0.282, 1.692 |
| **Target Altitude [km]** | H2 | 0.168 |
| **Day Number** | IDAY | 312 |
| **Latitude** | PARM1 | 43.041 |
| **Longitude** | PARM2 | 77.698 |
| **Time [UTC]** | TIME | 18.0 |
| **Starting Wavelength [um]** | V1 | 0.33 |
| **Ending Wavelength [um]** | V2 | 1.2 |
| **Wavelength Increment [um]** | DV | 0.001 |
| **FWHM [um]** | FWHM | 0.001 |

# REFERENCES

[1] Colomina, I. and Molina, P., "Unmanned aerial systems for photogrammetry and remote sensing: A review," *ISPRS Journal of Photogrammetry and Remote Sensing* **92**, 79 – 97 (2014).

[2] DeBiasio, M., Arnold, T., Leitner, R., McGunnigle, G., and Meester, R., "UAV-based environmental monitoring using multi-spectral imaging," *Proc.SPIE* **7668**, 7668 – 7668 – 7 (2010).

[3] Bondi, E., Salvaggio, C., Montanaro, M., and Gerace, A. D., "Calibration of UAS imagery inside and outside of shadows for improved vegetation index computation," in [*Autonomous Air and Ground Sensing Systems for Agricultural Optimization and Phenotyping*], **9866**, 98660J, International Society for Optics and Photonics (2016).

[4] Smith, G. M. and Milton, E. J., "The use of the empirical line method to calibrate remotely sensed data to reflectance," *International Journal of Remote Sensing* **20**(13), 2653–2662 (1999).

[5] Baugh, W. and Groeneveld, D., "Empirical proof of the empirical line," *International Journal of Remote Sensing* **29**(3), 665–672 (2008).

[6] Wang, C. and Myint, S. W., "A simplified empirical line method of radiometric calibration for small unmanned aircraft systems-based remote sensing," *IEEE Journal of Selected Topics in Applied Earth Observations and Remote Sensing* **8**, 1876–1885 (May 2015).

[7] MicaSense Incorporated, *MicaSense RedEdge 3 Multispectral Camera User Manual*.

[8] Schott, J., [*Remote Sensing: The Image Chain Approach*], Oxford University Press (2007).

[9] MicaSense Incorporated, "Image processing." https://github.com/micasense/imageprocessing (2017).

[10] Lekki, J., Anderson, R., Nguyen, Q.-V., Demers, J., Leshkevich, G., Flatico, J., and Kojima, J., "Development of hyperspectral remote sensing capability for the early detection and monitoring of harmful algal blooms (HABs) in the great lakes," in [*AIAA Infotech@ Aerospace Conference and AIAA Unmanned...Unlimited Conference*], 1978 (2013).

[11] Ortiz, J. D., Avouris, D., Schiller, S., Luvall, J. C., Lekki, J. D., Tokars, R. P., Anderson, R. C., Shuchman, R., Sayers, M., and Becker, R., "Intercomparison of approaches to the empirical line method for vicarious hyperspectral reflectance calibration," *Frontiers in Marine Science* **4**, 296 (2017).
[12] "MODTRAN." http://modtran.spectral.com/modtran_index. Accessed: 2017-03-27.
[13] McDowell, D., "Spectral distribution of skylight energy for two haze conditions," *Photogrammetric Eng* **40**, 569–571 (1974).
[14] "Spectral Vista Corporation, HR-1024i." https://www.spectravista.com/hr-1024i/. Accessed: 2017-03-27.
[15] Berk, A., Anderson, G., Acharya, P., Chetwynd, J., Bernstein, L., Shettle, E., Matthew, M., and Adler-Golden, S., "MODTRAN4 User's Manual," *Air Force Research Laboratory* **1** (1999).